\magnification=\magstep1
\hfuzz=12pt
\baselineskip=15pt
$ $
\vfill
\centerline{\bf Quantum controllers for quantum systems}
\bigskip
\centerline{Seth Lloyd}

\centerline{d'Arbeloff Laboratory}

\centerline{Mechanical Engineering}

\centerline{MIT 3-160, Cambridge, Mass. 02139}

\centerline{slloyd@mit.edu}

\bigskip\noindent{\bf Abstract:} Feedback control uses
sensors to get information about a system, a controller to
process that information, and actuators to supply controls.
In the conventional picture of quantum feedback and feedforward
control, sensors perform measurements on the system,
a classical controller processes the results of the measurements,
and actuators supply semiclassical potentials to
alter the behavior of the quantum system.  Since quantum measurement
inevitably disturbs the system, the conventional
picture portrays quantum feedback control as a stochastic process
during which the initial state of the system is destroyed.
This paper proposes an alternative method for quantum
feedback control, in which the sensors, controller, and
actuators are themselves quantum systems and interact
coherently with the system to be controlled: as a result,
the entire feedback loop is coherent.  Feedback
control by quantum controllers is not stochastic, preserves the
initial state of the controlled system, and can control
quantum systems in ways that are not possible using
conventional, incoherent feedback control.  In particular,
the target state to which the quantum controller drives the
system can be entangled with another quantum system.  This paper
investigates quantum controllers and states necessary and
sufficient conditions for a Hamiltonian quantum system to be
observable and controllable by a quantum controller.
\vfill\eject

Quantum control theory has a long history (1-4).  Experiments
on elementary particles, atoms, solid state systems,
and optics involve the systematic measurement and
manipulation of quantum systems.  In addition, control
theory has contributed significantly to the understanding
of fundamental aspects of quantum mechanics, including the
quantum Zeno effect (5-6), non-demolition measurements
(7-8), and stochastic quantization (9).
Formal aspects of quantum control theory were worked
out in the 1980's (1-3, 10), and
a host of practical applications of quantum control
have been realized in the 1990's as technologies have
matured (4,11-14).  Particularly significant
are experimental applications of optimal control theory
to quantum systems (10).  Many new technologies in
solid state, optical, and atomic physics owe their precision
and reliability to quantum control techniques.

The conventional method for controlling a quantum system
such as an atom using feedback is to make a measurement
on the system to determine its state, then to apply a semiclassical
potential such as a laser beam to guide the system to a
desired state.  This method may be termed semiclassical
control.  For example,
consider an atom in a quantum superposition of its ground and
excited state.  Stimulated emission can be used to measure
whether the atom is in the ground state or in the excited
state, leaving it in that state after the measurement.  If
the atom is found in the excited state, then a laser pulse
can be applied to the atom to `flip' it into its ground
state.  The net result of this feedback process is to
control the atom into its ground state.

{}From the perspective of control theory, semiclassical
control, though
effective, has several drawbacks.  First of all,
measuring a quantum system almost inevitably
disturbs it: even an unintrusive measurement
that leaves the system in the state in which it was measured
--- a so-called non-demolition measurement --- still alters
the state of the system prior to the measurement (7-8).
Suppose that the atom of the previous example is originally
in a coherent superposition of the ground and excited states.
After spontaneous emission determines whether the atom is in
its ground state or excited state the initial quantum coherence
between those states is irrevocably lost.
Secondly, semiclassical control is stochastic: as a result of
the measurement the system jumps to one state or another
probabilistically.

This paper reports on a new development in feedback and
feedforward control
of quantum systems.  As anyone who has built a PID
controller out of op-amps knows, sensors, controllers, and
actuators are dynamical systems in their own right.  In particular,
while the conventional view of quantum feedback control
looks at classical controllers interacting with quantum
systems through semiclassical sensors and actuators,
there is no reason why sensors, controllers, and actuators
should not themselves be quantum systems.  Quantum sensors
and actuators are familiar devices: a photon may get
information by scattering coherently off an electron
coherent fashion, and may act coherently on another electron
by scattering off it in turn.  A controller is
a dynamical system that
processes the information it gets from sensors and uses that
information to make decisions as to what actuators should
do.  Conventional information-processing devices such as
analog or digital computers do not behave in a quantum
mechanically coherent fashion.  Work on quantum
computers (15-18) indicates that both analog and digital quantum
devices can process information coherently, however,
and simple quantum logic devices have recently been
constructed (11-13).

As will be shown below, quantum controllers can perform
a number of tasks that semiclassical controllers cannot.  For
example, they can use coherent feedback to guide a
quantum system from an unknown state to a desired
state without destroying its initial state.
Although the state of the controller becomes
correlated with the state of the system during the control
process, no irreversible measurement is made, and the
initial state of the system is not lost.  In addition,
a quantum controller can drive a quantum system to a target
state that is entangled with another quantum system.
Entanglement is a non-local quantum phenomenon that cannot
be created by classical controllers.

Consider for example the problem of taking a quantum spin that is
originally in the state $|\psi\rangle= \alpha|\uparrow\rangle +
\beta|\downarrow\rangle$, where $\alpha$ and $\beta$ are unknown,
and putting it in the state $|\downarrow\rangle$.
In semiclassical control, the controller begins by making a
non-demolition measurement of the state of the spin (using,
say, a Stern-Gerlach apparatus),
giving $|\uparrow\rangle$ with probability $|\alpha|^2$ and
$|\downarrow\rangle$ with probability $|\beta|^2$.
The control algorithm is as follows:
if the result of the measurement is $|\downarrow\rangle$,
do nothing, while if the result of the measurement is
$|\uparrow\rangle$, put the spin in a static magnetic field
$B$ and apply an electromagnetic pulse with frequency
$\omega=2\mu B/\hbar$ to flip the spin (here $\mu$ is the spin's
magnetic dipole moment).
The spin is now in the state $|\downarrow\rangle$ as
desired.  The control process is stochastic, and although the
measurement reveals the state of the spin along some axis,
it destroys the original coherent superposition.
In the normal idealization of classical feedback
and feedforward control, sensors do not disturb the system
about which they get information.  In semiclassical control
of quantum systems, however, measurement introduces an
irreversible, stochastic disturbance.

To contrast to the semiclassical case,
consider a quantum controller consisting of a second
spin, initially in the state $|\downarrow\rangle'$, that
interacts with the first through the usual scalar
interaction term $\gamma \sigma_z \sigma'_z$ (19-20)
so that the Hamiltonian for the two spins
is $(\hbar/2)(\omega \sigma_z + \omega' \sigma'_z +
\gamma \sigma_z \sigma'_z)$, where $\omega'=2\mu' B/\hbar
\neq \omega$ is the resonant frequency of the second spin.
The quantum sensors and actuators operate by enhancing
the spin-spin interaction using conventional magnetic
resonance techniques.  For example, applying a pulse with
frequency $\omega'+\gamma$ coherently flips the second spin if and
only if spin 1 is in the state $|\uparrow\rangle$.  The two
spins are now in the state
$\alpha |\uparrow\rangle
|\uparrow\rangle' + \beta |\downarrow\rangle
|\downarrow\rangle'$.
Clearly, the second spin has become
correlated with the first spin in the sense that measuring
the state of the second spin would reveal the state of the
first spin.  The state of the first spin has also been
disturbed: while initially it was described by the pure-state
density matrix
$$\rho = |\psi\rangle\langle\psi|=
\alpha\bar\alpha |\uparrow\rangle \langle \uparrow|
+\alpha\bar\beta |\uparrow\rangle \langle \downarrow|
+\beta\bar\alpha|\downarrow\rangle \langle \uparrow|
+\beta\bar\beta |\downarrow\rangle \langle \downarrow|,$$
\noindent it is now described by the mixed-state density matrix
$\rho' =
\alpha\bar\alpha |\uparrow\rangle \langle \uparrow|
+\beta\bar\beta |\downarrow\rangle \langle \downarrow|$.
No irreversible measurement has taken place, however.  The
disturbance can be removed and the correlation undone by
applying a second pulse with the same frequency to flip
the second spin back again, returning both spins to their
initial state.  With quantum controllers,
in contrast with semiclassical controllers,
the disturbance introduced by the sensors
is reversible and can be undone by the actuators.

A second coherent interaction between the two spins now
controls the spin coherently to the state $|\downarrow\rangle$:
simply apply to the system in state $\alpha |\uparrow\rangle
|\uparrow\rangle' + \beta |\downarrow\rangle
|\downarrow\rangle'$
a pulse with frequency $\omega+\gamma$ to flip
the first spin if and only if the second spin is up.
The state of the two spins is now $|\downarrow\rangle(\alpha
|\uparrow\rangle' + \beta |\downarrow\rangle')$.  That is,
not only has coherent quantum feedback put the first
spin in the state $|\downarrow\rangle$, it has coherently
put the second spin in the initial state of the first spin.
No stochastic operation has taken place, and the initial
state of the controlled spin has not been destroyed: rather,
it has been coherently transferred to the state of the
controller.

Finally, suppose that the goal of the control process is to
put the spin in a state
$(1/{\sqrt 2}) (|\uparrow\rangle
|\uparrow\rangle'' + |\downarrow\rangle|\downarrow\rangle''
)$ where $|\uparrow\rangle''$ and $|\downarrow\rangle''$ are
states of a third spin that is not acted on by the
controller.  Such states are called {\it entangled} because
they exhibit peculiar non-local quantum effects the best
known of which is the Einstein-Podolsky-Rosen (EPR) effect (21).
Creating and controlling entangled states is a crucial part
of new quantum technologies such as quantum cryptography,
quantum computation, and teleportation (15-18,22).  It is
easily verified that the procedure given in the previous two
paragraphs accomplishes the goal of producing this
entanglement provided that the initial target state stored
in the controller is itself entangled with the third spin.
The initial state of the three spins is then
$$(\alpha|\uparrow\rangle+\beta|\downarrow\rangle) (1/{\sqrt
2}) ( |\uparrow\rangle'
|\uparrow\rangle'' + |\downarrow\rangle'|\downarrow\rangle''
)$$
\noindent and the final state is
$$(1/{\sqrt
2}) ( |\uparrow\rangle
|\uparrow\rangle'' + |\downarrow\rangle|\downarrow\rangle'')
(\alpha|\uparrow\rangle'+\beta|\downarrow\rangle'), $$
\noindent in which the first spin is entangled with the
third spin.  The controller never acts on the third spin,
which may be spatially distant from the first two.
A semiclassical controller cannot drive the
system to such an entangled target state without acting on
the third spin directly.

This example can be generalized to the control of an
arbitrary quantum system.  Two important questions for any
control method (23) are controllability: can a system be guided
to a desired state? --- and observability: can the
sensors determine the state of the system?
It is clear from the example above that controllability and
observability take on different guises when the controller
is semiclassical and when it is quantum-mechanical.
We now present
necessary and sufficient conditions for controllability and
observability of Hamiltonian quantum systems using both semiclassical and
quantum controllers.

First, consider semiclassical control of quantum systems.
A semiclassical controller applies time-dependent
potentials $\sum_i \gamma_i(t){\cal O}_i$ to the system.
Controllability is the problem of
taking a quantum system from some initial
state to a desired final state.
A quantum system will be said to be open-loop controllable
if the potentials $\gamma_i(t)$ can be chosen to take
the system from an arbitrary known
initial state $|\psi\rangle$ to a
desired final state $|\psi_d\rangle$.  This form of
controllability is called open-loop because the initial
state of the system is assumed to be known, and no
measurement is made on the system.
The problem of open-loop controllability has
an elegant geometric solution (10,17,24,25):

\bigskip\noindent{\it (1) Semiclassical controllability:
open-loop case.}

A quantum system with Hamiltonian $H$ is open-loop
controllable by a semiclassical controller
if and only if the algebra ${\cal A}$ generated from
$\{H, {\cal O}_i\}$ by commutation is the full algebra of
Hermitian operators for the system.
\bigskip

\noindent The spin in the example above is open-loop
controllable by a semiclassical controller
since NMR methods allow it to be taken from any
given state to any desired state: the algebra generated by
the Hamiltonian corresponding to the static field,
$B\sigma_z$, and the applied Hamiltonian, $B_x\sigma_x{\rm sin}
\omega t$, can easily be seen to generate the full
algebra of $SU(2)$ by commutation.
Result (1) is a quantum analog of the geometric
theory of classical nonholonomic control theory (26).
A familiar example of a classical nonholonomic control
problem is parallel parking: a car cannot be driven
sideways directly, but can still be parked by edging
first in one direction then in another.  In the quantum case,
the algebra ${\cal A}$ determines what set of states can be
reached be edging the quantum system first in one direction, then
in another, a method that can be called `parking Schr\"odinger's
car.'

Since the action of the semi-classical controller is
Hamiltonian, open-loop semiclassical control is limited to
taking a known pure state to a desired pure state.
To extend this controllability result to unknown mixed initial states
$\rho$ and to mixed final states $\rho_d$, we must introduce
closed-loop control.  Suppose that the controller can make
measurements on $S$ (for the sake of simplicity, assume
that these measurements are non-demolition measurements)
corresponding to a finite set of Hermitian observables
$\{{\cal M}_j\}$ and then apply potentials
$\sum_i \gamma_i(m_j, t) {\cal O}_i$ that depend on the results
$m_j$ of the measurements.  A quantum system $S$ will be said to be
closed-loop controllable if and only if a closed-loop controller
can take $S$ from an arbitrary unknown
initial state $\rho$ to any desired final state $\rho_d$.
We then have the following result:

\bigskip\noindent{\it (2) Semiclassical controllability:
closed-loop case.}

A quantum system with Hamiltonian $H$ is closed-loop
controllable by a semiclassical controller if and only if
(i) at least one of the ${\cal M}_j \neq I$ --- that is, the
controller can make some nontrivial measurement on the system---
and (ii) the algebra generated by $\{ H, {\cal O}_i \}$
is the full algebra of Hermitian operators for the system.

\bigskip\noindent For example, the spin above is clearly
closed-loop controllable by the semiclassical technique
described.  The proof of (2) is
somewhat detailed (27), and follows
from results presented in reference (18).  The `if' part
follows because even when one can only make an non-demolition
measurement of a single bit of information, the open-loop
controllability of the system allows that bit to correspond
to projections onto arbitrary subspaces; repeated measurements
then allow the value of any operator to be determined and
the system to be guided to a desired state.  The `only if'
part follows because if the system is not open-loop controllable,
then the set of states that can be reached conditioned on
the results of measurements is of lower dimension than the
Hilbert space of the system.

The close relationship between open- and closed-loop
controllability for quantum systems has implications for the
related notion of observability.  The classical definition
of observability must be somewhat altered for quantum
systems since the irreversible disturbance introduced
by measurement implies that no procedure can reveal
the precise initial state of a quantum system.
Accordingly, a quantum system will be
called observable by a semiclassical controller if
the proper sequence of controls and measurements
can be used to observe any desired feature of the initial
state of the system.  Specifically, the system is observable
if the controller can make a measurement that reveals the
projection of the original state along any desired set of
orthogonal axes in Hilbert space.  Result (2) immediately implies:

\bigskip\noindent{\it (3) Semiclassical observability.}

A Hamiltonian
quantum system is observable by a semiclassical controller
if and only if it is closed-loop controllable.

\bigskip\noindent In the example above, NMR techniques, together with
the ability to measure the component of spin along the $z$ axis,
clearly allow one to measure the spin along any axis.  In
addition, if one can manipulate the spin so as to measure it
along any axis, then one can also manipulate it sufficiently
to control its state to any desired state, conditioned on
the result of the measurement.

Now let's turn to the fully quantum controller.
Assume that the system interacts
with a quantum controller via
a coherent interaction $\sum_i\gamma_i(t) {\cal O}_i {\cal O}'_i$,
where ${\cal O}_i, {\cal O}'_i$ are Hermitian operators
acting on system and controller respectively and the $\gamma_i(t)$
are coupling constants that can be `turned on' and `turned off'
to make the system and controller interact.  We assume that at
least one ${\cal O}_i{\cal O}'_i$ pair is nontrivial in the
sense that neither ${\cal O}_i$ nor ${\cal O}'_i$ is the identity
operator: otherwise this case reduces to the semiclassical
case above.  In analog with ordinary digital control,
where the controller is a digital computer with
programmable dynamics, the controller is assumed to have
access to an arbitrarily large Hilbert space and
the dynamics of the controller are assumed to be
programmable to any desired dynamics.

For the fully quantum controller, there is no distinction between
open- and closed-loop control: an interaction that can
function as an actuator can also function as a sensor, and
{\it vice versa}.  A quantum system will be said to be
controllable by a quantum controller if
there is some initial state for the controller
(possibly entangled with
the state of another quantum system), a
dynamics for the controller and a schedule
of interactions $\gamma_i(t)$ that takes the system from
some initial state $\rho$ to a desired final state
$\rho_d$ which can also be entangled with another quantum
system.   We then have,

\bigskip\noindent{\it (4) Quantum controllability.}

A quantum system with Hamiltonian $H$ is
controllable by a quantum controller
if and only if the algebra ${\cal A}$ generated from
$\{H, {\cal O}_i\}$ by commutation is the full algebra of
Hermitian operators for the system.
\bigskip

\noindent Result (4) for quantum controllers conflates
results (1-2) for semiclassical controllers.  Essentially,
the equivalence between quantum sensors and quantum actuators
makes quantum control intrinisically closed-loop.
For example, the one-spin quantum controller above is
clearly capable of controlling the other spin to any desired
state, entangled or not.

Just as in the semiclassical case,
care must be taken in defining observability for quantum
controllers: the controller is not a classical device that makes
measurements on the system, but a quantum system in its own
right that becomes correlated with the system.
No irreversible measurement ever takes place.  A quantum
system will be said to be observable by a quantum controller
if the initial state of the system, together with all its
entanglements with any other quantum systems, can be
transferred to an analogous state of the controller.
The controller can then use this transferred state as the
target state to which to control some other quantum system.
This fundamentally quantum definition of observability is
the natural converse to the quantum definition of
controllability in (4).  Given results (1-4), the following
result should come as no surprise:

\bigskip\noindent{\it (5) Quantum observability.}

A Hamiltonian quantum system is observable by a quantum
controller if and only if it is controllable by the
controller.

\bigskip\noindent As the example of the three spins shows,
an interaction with a quantum controller that puts a
a Hamiltonian system in a desired state necessarily
transfers the initial state or the system, together with its
entanglements, to an analogous final state of the
controller (28).

In conclusion, this paper explored the properties of
semiclassical and quantum controllers and gave necessary and
sufficient conditions for quantum controllability and
observability, which turned out to be equivalent for Hamiltonian
quantum systems.  Quantum
controllers are likely to play a key role in the development
of quantum technologies such as quantum computation and
quantum communications.  Although the potential experimental
realizations of quantum controllers discussed here
were based on nuclear magnetic resonance, this paper's
results could also be
realized using quantum logic devices such as ion traps
(12), high-Q cavities in quantum optics
(11,13-14), and quantum dots (29).
Quantum controllers could have application to a variety
of problems, including problems with classical analogs
such as trajectory control, and problems with no
classical analog such as preventing decoherence.
As the theory of quantum error correction shows (30),
strategies for disturbance rejection are harder
to devise for quantum systems than for classical.
A particularly important open question is whether the
controllability and observability results reported here for
Hamiltonian quantum systems can be extended to
open quantum systems.

\vfill
\noindent{\it Acknowledgements:} This work was supported by
grants from ONR and from DARPA/ARO under the Quantum
Information and Computation Initiative (QUIC).  The author
would like to acknowledge helpful discussions with
I. Chuang, H. Mabuchi, D. Rowell, H.A. Rabitz, and J.J. Slotine.
\eject
\centerline{\bf References}
\bigskip
\noindent(1) A. Blaquiere, S. Diner, G. Lochak, eds., {\it
Information Complexity and Control in Quantum Physics}
(Springer-Verlag, New York, 1987).  This volume constitutes
the Proceedings of the 4th International Seminar on Mathematical
Theory of Dynamical Systems and Microphysics, Udine,
September 4-13, 1985, and contains multiple references to
work in quantum control theory before 1985.

\noindent(2) A. Blaquiere, {\it Modeling and Control of
Systems in Engineering, Quantum Mechanics, Economics and
Biosciences} (Springer-Verlag, New York, 1989).  This volume
constitutes the Proceedings of the Bellman Continuum
Workshop, Sophia Antipolis, June 13-14, 1988.

\noindent(3) A.G. Butkovskiy, Yu.I. Samoilenko, {\it Control
of Quantum-Mechanical Processes and Systems} (Kluwer
Academic, Dordrecht, 1990).

\noindent(4) H. Ezawa, Y. Murayama, eds., {\it Quantum
Control and Measurement} (North-Holland, Amsterdam, 1993).
Proceedings of the ISQM Satellite Workshop ARL, Hitachi,
Hatoyama, Saitama, August 28-29, 1992.

\noindent(5) C.B. Chiu, E.C.G. Sudarshan, B. Misra, {\it
Phys. Rev. D} {\bf 16}, 520 (1977).

\noindent(6) A. Peres, {\it Am. J. Phys.} {\bf 48}, 931
(1980); in reference (1), 235.

\noindent(7) V.B. Braginsky, Y.I. Yorontsov, K.S. Thorne,
{\it Science} {\bf 209}, 547 (1980).

\noindent(8) C.M. Caves, K.S. Thorne, R.W.P. Drever, V.D.
Sandberg, M. Zimmerman, {\it Rev. Mod. Phys.} {\bf 52}, 341
(1980).

\noindent(9) S. Mitter, in reference (2), 151.

\noindent(10) A. Peirce, M. Dahleh, H. Rabitz,
{\it Phys. Rev. A} {\bf 37}, 4950
(1988); {\it ibid} {\bf 42}, 1065 (1990); R.S. Judson, H.
Rabitz {\it Phys. Rev. Lett.} {\bf 68}, 1500 (1992); W.S.
Warren, H. Rabitz, M. Dahleh, {\it Science} {\bf 259}, 1581 (1993);
V. Ramakrishna, M.V. Salapaka, M. Dahleh, H. Rabitz, A. Peirce,
{\it Phys. Rev. A} {\bf 51}, 960 (1995).

\noindent(11) Q.A. Turchette, C.J. Hood, W. Lange, H.
Mabuchi, H.J. Kimble, {\it Phys. Rev. Lett.} {\bf 75}, 4710
(1995).  H. Mabuchi, H.J. Kimble, {\it Opt. Lett.} {\bf 19},
749 (1993); A.S. Parkins, P. Marte, P. Zoller, H.J. Kimble,
{\it Phys. Rev. Lett.} {\bf 71}, 3095 (1993).

\noindent(12) C. Monroe, D.M. Meekhof, B.E. King, W.M.
Itano, D.J. Wineland, {\it Phys. Rev. Lett.} {\bf 75}, 4714
(1995).

\noindent(13) M. Brune {\it et al.}, {\it Phys. Rev. Lett.}
{\bf 72}, 3339 (1984); P. Domokos, J.M. Raimond, M. Brune,
S. Haroche, {\it Phys. Rev. A} {\bf 52}, 3554 (1995).

\noindent(14) H. Walther, in reference (4), 113.

\noindent(15) For a review of quantum computing, see
D. Divincenzo, {\it Science} {\bf 270}, 255 (1995).

\noindent(16) S. Lloyd, {\it Sci. Am.} {\bf 273}, 140
(1995).

\noindent(17) S. Lloyd, {\it Science} {\bf 273}, 1073
(1996).

\noindent(18) A. Ekert, to be published.

\noindent(19) C.P. Slichter, {\it Principles of Magnetic
Resonance}, third edition, (Springer-Verlag, New York,
1990).

\noindent(20) O.R. Ernst, G. Bodenhausen, A. Wokaun, {\it
Principles of Nuclear Magnetic Resonance in One and Two
Dimensions} (Oxford University Press, Oxford, 1987).

\noindent(21) A. Einstein, B. Podolsky, N. Rosen, {\it Phys.
Rev.} {\bf 47}, 777 (1935).

\noindent(22) C.H. Bennett, {\it Physics Today} {\bf 48}, 24
(1995).

\noindent(23) D.G. Luenberger, {\it Introduction to Dynamic
Systems} (Wiley, New York, 1979).

\noindent(24) T.J. Tarn, J.W. Clark, G.M. Huang, in
reference (2), p. 161.

\noindent(25) Strictly speaking, results (1-5) hold only for
systems with finite energy confined to a finite region of space.

\noindent(26) R.W. Brockett, R.S. Millman, H.J. Sussman,
eds., {\it Differential Geometric Control Theory}
(Birkhauser, Boston, 1983).  Z. Li, J.F. Canney, eds., {\it
Nonholonomic Motion Planning} (Kluwer Academic, Boston,
1993).

\noindent(27) S. Lloyd, to be published.

\noindent(28) The transfer of state from system to
controller also occurs in the closed-loop control of
classical Hamiltonian systems.

\noindent(29) C.B. Murray, C.R. Kagan, M.G. Bawendi, {\it
Science} {\bf 270}, 1335 (1995).

\noindent(30) P.W. Shor, {\it Phys. Rev. A} {\bf 52}, 2493 (1995).
A.R. Calderbank and P.W. Shor, {\it Phys. Rev. A} {\bf 54}, 1098
(1996).  A.M. Steane, {\it Phys. Rev. Lett.} {\bf 77}, 793 (1996).
R. Laflamme, C. Miquel, J.P. Paz, W.H. Zurek, {\it Phys. Rev.
Lett.} {\bf 77}, 198 (1996).  C.H. Bennett, D.P. DiVincenzo,
J.A. Smolin, W.K. Wootters, {\it Phys. Rev. A} {\bf 54}, 3824 (1996).

\vfill\eject\end